\documentclass{elsart}

\usepackage{graphicx}
\usepackage{amssymb}

\begin{document}
%%%%%%%%%%%%%%%%%%%%%%TITLE%%%%%%%%%%%%%%%%%%%%%%%%%
\begin{frontmatter}

\title{MONTE CARLO SIMULATION AND STATISTICAL ANALYSIS OF GENETIC INFORMATION CODING}
\author{E. Gultepe\corauthref{Northeastern}}
\author{M.~L. Kurnaz\corauthref{BU}}
\corauth[Northeastern]{Present Address: Northeastern University}
\address{Department of Physics, Bogazici University, 34342
Bebek Istanbul}
\ead{kurnaz@boun.edu.tr}
\corauth[BU]{Corresponding Author}

%%%%%%%%%%%%%%%%%%%%%ABSTRACT%%%%%%%%%%%%%%%%%%%%%%%
\begin{abstract}
The rules that specify how the information contained in DNA codes
amino acids, is called ``the genetic code". Using a simplified
version of the Penna nodel, we are using computer simulations to
investigate the importance of the genetic code and the number of
amino acids in Nature on population dynamics. We find that the 
genetic code is not a random pairing of codons to amino acids and 
the number of amino acids in Nature is an optimum under mutations.
\end{abstract}

\end{frontmatter}

\section{INTRODUCTION\protect\\ }
\label{sec:level1}
In general population dynamics is a matter of interest for biologists, 
however it has attracted the attention of physicists since it is a 
subject very closely related to statistical mechanics. Investigation 
of population dynamics in Nature is not a simple task because to 
get any idea about the dynamics of population growth, one has to 
consider many generations of a population. Even if one can find 
fast-reproducing species like the fruit fly, checking all individuals in
such a population is not an easy task either. Therefore, modelling
with computers has lots of advantages such as considerably small
time consumption and simplicity in population monitoring.

The most successful computational model for age-structured
populations is the Penna model \cite{Penna}. In this model, 
individuals are represented by bit-strings which are 32 bits long 
and are initially set to zero. Each bit represents a given age: 
as the individual gets older we move down on the bit-string.
Bits which are set to zero represent that no bad
mutations is stored at that age. However, if a bit is set to one,
it means that the individual suffers a disease at that age and its
probability of staying alive is decreased.

The Penna model has been successfully used to investigate the 
advantages of sexual reproduction over asexual reproduction
\cite{Stauffer13580}\cite{Stauffer13600}\cite{Stauffer13570}\cite{Sousa680}\cite{Tuzel2770}\cite{Tuzel21970}\cite{Orcal3410}, 
certain features of ecology \cite{Penna2510} and population dynamics 
\cite{Penna13590}\cite{Penna13610}\cite{Huang3140}.

To investigate the importance of \textit{the genetic code} and 
number of amino acids in population dynamics we have constructed
a model based on the Penna model.

The genetic information about the individuals is stored in the DNA.
DNA is made up of different monomers. Each monomer, nucleotide, 
in the chain carries a heterocyclic base. In DNA, these bases 
are adenine (A), guanine (G), cytosine (C) and thymine (T). 
Proteins are synthesized from amino acids using the information 
stored in the DNA. As there are four bases in the DNA, and 20 amino
acids used in proteins, during protein synthesis there is not a
one-to-one correspondence between the nucleotides in the DNA and the
amino acids in a protein. Rather, the linear sequence of bases
which constitutes the protein-coding information is "read" by the
cell in blocks of three nucleotide residues, or codons, each of
which specifies a different amino acid. If we consider a
nucleotide on the DNA to be a letter in a four-letter alphabet,
codons can be thought as words with three letters. Hence, there
are sixty four words to code the twenty different amino acids. The
set of rules that specifies which nucleic acid codons correspond
to which amino acid is known as the genetic code.

\section{COMPUTATIONAL METHOD\protect\\ }
\label{sec:level2}

If there is a mutation on a gene which causes a change in the
amino acid chain, we will think that the organism may not be able
to build the protein which may be crucial for the organism. If so;
it will not function properly or it may simply die; hence it is
simplistic yet reasonable to represent the organism by a single
gene.

In our model, to represent a whole individual, we took a real gene
from Nature ``human cytokine" (LD78 Homo sapiens blood lymphocyte
gene on the DNA 17$^{th}$ chromosome) \cite{gene}. This gene is
necessary for activating lymphocytes; therefore if it is missing
the human body cannot perform immune responses.

If all other effects (aging, food restriction, illness etc.) are
neglected, mutation will be the only possible cause of death.
Also, in our simplistic model reproduction is not included, 
therefore we have a population which can only decrease as a 
result of mutations. We use this model to investigate the 
effects of mutations to population decrease.

A mutation in our model is a process acting at each site 
independently. We disregard more complicated processes such as
deletions or insertions, and we only look at single nucleotide 
replacements by another nucleotide in the gene. Normally the 
rates for these replacements depend on the two nucleotides being 
interchanged. The simplest approach to the problem is to take all
mutation rates to be equal, known as the Jukes-Cantor mutation scheme
\cite{Jukes21920}.

The mutation is taken to be deleterious if it causes a change in 
the amino acid chain; and not all the mutations kill the individual. 
A real gene is composed of two different parts: a coding portion and 
a noncoding portion. The coding part, exon, is responsible for coding 
for proteins whereas the rest, intron, does not code for a protein and 
the purpose of this part is not clearly understood yet. If a mutation
takes place on intron part, it is considered to be simply harmless
but if it takes place on exon part, it is usually harmful, but there 
is still a chance: The interchanged codon may still code the same amino 
acid since more than one codon can code one amino acid in nature.

To be more explicit, the codons AAA and AAG code the same amino
acid, ``lysine''; hence if AAA turns into AAG as a result of a
mutation the amino acid will not change and the protein can be
constructed safely. However; if AAA turns into AGA, which codes
the amino acid ``arginine'', the amino acid chain will change and
we assume that the protein can not build up, which means the
represented organism will die.

There can be a mutation which converts AAA to AAX where X $\neq$
{A, G, C, T}; then the individual dies automatically. As a model,
we are looking at a simpler case where a mutation changes A to one
of G, C, T not X.

Since reproduction is not included in the model, the population
can only diminish. The decrease in population can be found by
calculating the probability of a deleterious mutation. The
probability of the mutation changing the amino acid depends on 
the codon; so one needs to find the probability of hitting each 
different codon type. First, the probability of hiting a codon
type ($P_{\alpha}$) is calculated as the ratio of the number 
of codons of that type in the gene ($N_{\alpha}$) to total 
number of codons. Then we need to exclude the mutations that
do not cause a change in the amino acid and calculate the 
probability of a change occurring in the amino acid caused by
a change in one nucleotide ($P(d/{\alpha}$)) is calculated. 

As an example; only two codons code the amino acid ``lysine'': 
AAA and  AAG. In the exon part of human cytokine gene, there are 
only three ``AAA'' codons and the total number of codons in the gene 
is 207, hence the probability of the mutation hitting an ``AAA'' 
codon is simply $P_{\alpha} = 3\div207 = 0.0145$. By a point 
mutation to ``AAA" we can have 9 different codons (AAC, AAG, AAU, 
ACA, AGA, AUA, CAA, GAA, UAA). One of these codons still codes 
the same amino acid (AAG). Therefore the probability of deleterious
mutation ($P(d/{\alpha})$) is 8/9 for ``AAA'' in the human cytokine
gene.

Next, we need to calculate the probability of hitting the exon part 
of the gene as the ratio of the exon part to the total gene. In the 
human cytokine gene, there are 621 nucleotides in exon part and 
1447 ones in intron part:
\begin{equation}
P(hitting \: exon) = \frac{621}{2068}= 0.3032\label{}
\end{equation}

\noindent Hence; the probability of having a deleterious mutation
for all of the gene is simply:
\begin{equation}
P(deleterious) = P(hitting \, exon) \sum_{\alpha = 1}^{64} [
P_{\alpha} P(d/\alpha)]  = 0.2344 \label{}
\end{equation}

\noindent The survival probability can be calculated by:

\begin{equation}
P(surviving) = 1- P(deleterious) = 0.7656 \label{}
\end{equation}

\noindent If we take a population of $N_0$ gene (individuals),
after n mutations, to the first approximation, the number of 
surviving individuals is given by:

\begin{equation}
N_n  \approx N_0P(surviving)^n \label{}
\end{equation}

\noindent Hence, we obtain the ``probability of survival" with
the slope of the number of surviving individuals versus 
time graph:

\begin{equation}
slope \approx ln[P(surviving)] = -0.2670 \label{slope}
\end{equation}

During this calculation we used a simple assumption that after each
timestep the genome remain the same as the wildtype. A mutation may 
result in a different nucleotide sequence, but if this sequence codes 
the same amino acid, we assume that this mutation has never happened 
and go to the next stage. Hence, all alive individuals can be represented
by the same array, wildtype, as in the calculations. However, in the 
less likely event of a harmless mutation the number of the codons of
each type changes which will in turn slightly change the probabilities. 
We have designed a test simulation where after each mutation and deletion
of the individual, we have set all the sequence back to the wildtype. As 
this simulation gives the same results (within the error bars) as the
original case, we have used the modified sequence in the later stages.

\section{SIMULATION\protect\\ }
\label{sec:level3}

In the simulation, an individual (a gene) is represented by an array 
which contains 0, 1, 2, and 3's instead of the nucleotides Adenine (A), 
Guanine (G), Cytosine (C) and Thymine (Uracil (U)) respectively and 
also a sign bit which shows if the gene has a deleterious mutation (1) 
or not (0).

\begin{figure}[!]
\begin{center}
\includegraphics[width=14cm]{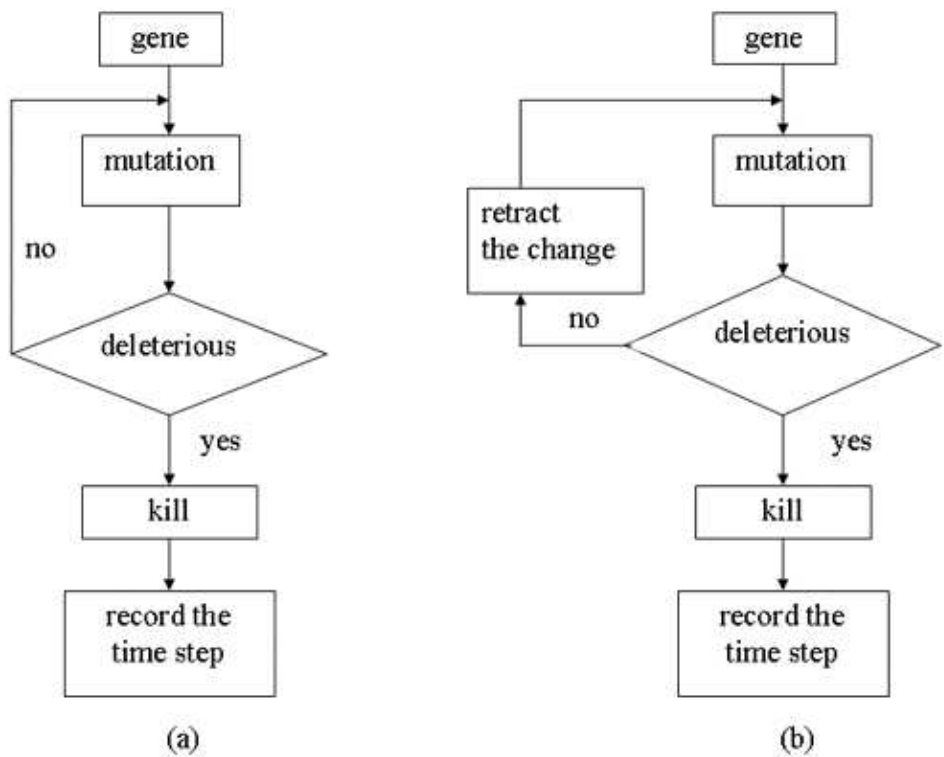}
\caption{a.) Flowchart of the simulation
b.) Flowchart of the control simulation}
\label{flowchart}
\end{center}
\end{figure}

In each ``cycle", each individual has to go through one mutation event, 
then it is determined whether or not the individual should die. In the 
mutation event; the place of mutation and the mutant nucleotide is 
determined randomly. If the nucleotide is not in the exon part, the 
sign bit remains 0. Otherwise, the changed codon is checked for the 
amino acid which it codes. If it is coding the same amino acid, the 
protein can still be built, therefore the sign bit is not changed and 
the individual survives. However, if the amino acid is changed then the
sign bit becomes '1'  that means this individual will be deleted. Deletion 
time is recorded for each individual [Fig.\ref{flowchart} (a)]. In the 
control simulation, if the mutation is harmless, modification of the gene 
will be recovered [Fig.\ref{flowchart}(b)].

After $N$ individuals, the number of surviving individuals in each time 
step is calculated. Since the probability of mutation is independent
of the number of individuals, this number also gives us the population 
size. Hence, we have an exponential population decay and the exponent 
depends on the probability of surviving ($P(surviving)$). Logarithm of 
the population is fitted to a straight line and the slope of the line 
is calculated.

We run all simulations ten times. The average of the slopes of the 
control simulations  $-0.266 \pm 0.001$, which is noticeably close to 
the slope derived from calculations. After the control, we run the 
simulation using genetic code of Nature. One example of such runs is 
shown in Figure \ref{slope}. The average of slopes for Nature's 
simulation is $-0.266 \pm 0.001$.\\

\begin{figure}[!]
\begin{center}
\includegraphics[width=14cm]{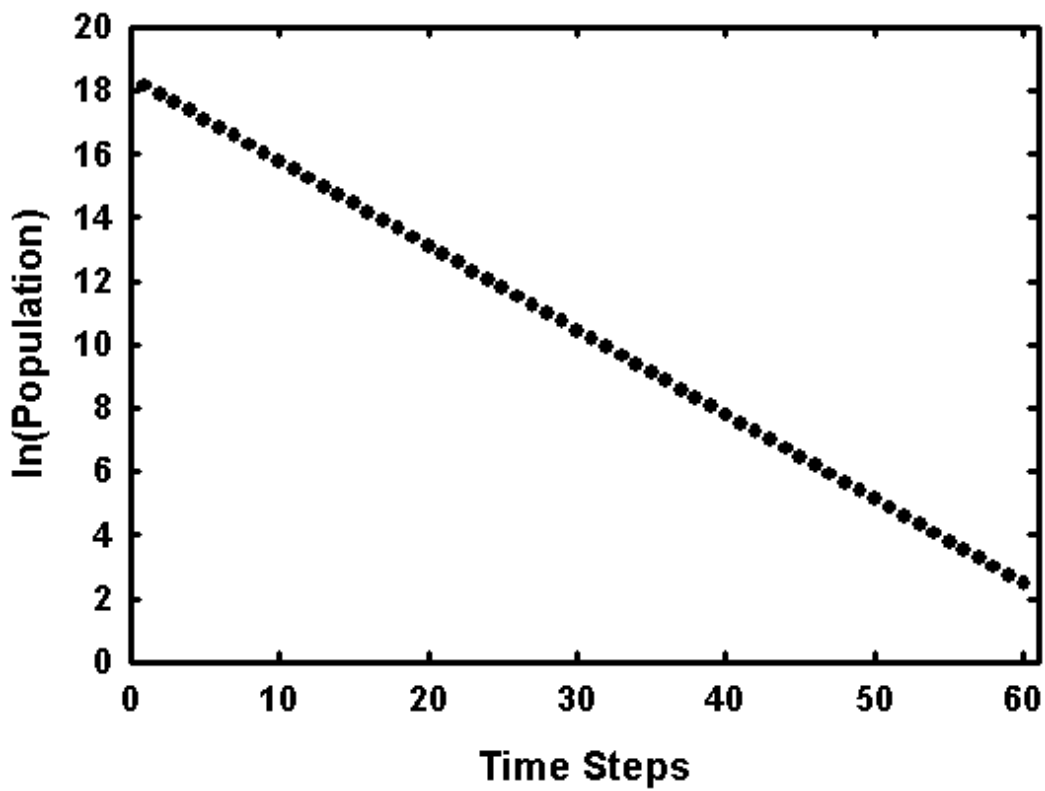}
\caption {Population decreasing: one of the simulations of the
amino acids table of Nature}\label{slope}
\end{center}
\end{figure}

\section{ARTIFICIAL TABLES\protect\\ }
\label{sec:level4}

With a few exceptions, twenty different kinds of amino acids 
are used to build the proteins. Even though in some rare
cases certain organisms use selenocysteine and pyrrolysine,
in Nature, the majority uses the same table. Recently a team of
investigators at the Scripps Research Institute modified a 
form of the bacterium Escherichia coli to use a 22-amino acid
genetic code instead of 20 \cite{Anderson21930}. They have
engineered the modified form of E. Coli to make myoglobin
proteins with 22 amino acids, using the unnatural amino acids
O-methyl-L-tyrosine and L-homoglutamine in addition to the 
naturally occurring 20. This work opens up the possibility
that the same procedure can be used to expand the amino acid
family even further. So, the question is why did life stop 
with twenty amino acids? To investigate the importance 
of the number of amino acids, we create amino acid tables 
based on Nature's table but with different amino acid numbers 
and we use them in the simulation.

If we change the number of amino acids in the genetic code, it
means that we change the amount of information in the genome.
Hence, to conserve the information, the genome length needs to be
adjusted also. Moreover, if we want our simulation to represent
Nature, the process of extending or shortening the amino acid
table needs to obey some rules of biochemistry.

The twenty amino acids contain with their twenty different side
chains of different chemical properties. This allows proteins to have
such a great variety of structures and properties. There are
several classes of side chains, grouped by their dominant chemical
features. While developing tables, we try to make them fit the
natural structure of amino acids obeying the classification in
\cite{Matthews}.

We have tried to use amino acid tables with more and with less number 
of amino acids. To shorten the amino acid tables, we first randomly 
choose which amino acid will be removed from the table. The random 
choice is made such that no two amino acids are removed from the 
same group (as long as the number of removed amino acids is less than 
the number of amino acids groups).

For example, in the table which has 18 amino acids Glutamine and 
Isoleucine are removed. Glutamine is in the acidic group and its 
frequency of occurrence is $3.9 \%$. The codons which code Glutamine 
formerly (CAA, CAG) will code Glutamic Acid which is also in acidic 
group and has the frequency of $6.2 \%$. Isoleucine is in the aliphatic 
group and its frequency is $4.6 \%$. The codons which code Isoleucine 
formerly (AUU, AUC, AUA) will code Glycine which is also in aliphatic 
group and has the frequency of $7.5 \%$. 

To conserve the information content of the gene, the gene should be 
lengthened. As an example, by changing the codons CAA and CAG from
Glutamine to Glutamic Acid we have lost the information carried by 
Glutamine. Now we take Asparagine and Aspartic Acid, which are also
in the acidic group, and insert them where Glutamine was originally.
The same procedure was repeated to decrease the number of amino acids
to 16 and then to 14.

To extend the amino acid table, first we determine which consecutive 
amino acid pair in the gene has the highest frequency. To do this, we 
calculate the number of occurrence of pairs and construct the pair matrix. 
Then, each frequent pair is replaced by a new amino acid. For example,
Leucine - Leucine pair has the highest frequency and they will be replaced
by the new amino acid called New1. Similarly Threonine - Serine pairs are 
replaced by New2. Leucine is from the aliphatic group, and New1 is 
constructed by dividing Alanine, which is also from the aliphatics group
and is represented by the highest number of codons. Now the codons GCU 
and GCC still code Alanine but the remaining (GCA, GCG) will code New1. 
Similary, New2 is formed by dividing Serine: the codons UCU, UCC, UCA, 
and UCG code still Serine but the others (AGU and AGG) code New2. The 
tables with 24 amino acid , with 26 amino acids and with 28 amino acids
are constructed just the same way.

As control cases we have also done simulations with different amino acid 
tables, both increased and decreased number of amino acids, where we 
neglected the conservation of information and kept the genome length
constant. As expected, the results of these simulations were very close
(within error bars) to the results of the original table.

Biologists have also been trying to find simplified amino acid alphabets.
One of these methods is the MJ matrix constructed using Wang and Wang's 
method \cite{Wang21950} which is based on Miyazawa-Jernigan's (MJ) residue -
residue potentials \cite{Miyazawa21960}. Their reduction algorithm, which
connects different representations of a protein, is generally based on
the idea that the amino acids can be distributed into different groups,
with different interactions. The interactions between amino acids of two 
different groups should have similar characteristics for a successful
reduction.

Another method is the BLOSUM50 matrix, built using Murphy, Wallqvist and
Levy's method \cite{Murphy21940} derived by Henikoff and Henikoff 
\cite{Henikoff7460}. Their reduction scheme is based on the analysis of
correlations among similarity matrix elements used for sequence alignment.
We have constructed reduced amino acid tables using both the MJ matrix and
BLOSUM50 matrix methods.

\section{CONCLUSION\protect\\ }
\label{sec:level6}

In this paper, we developed a computer simulation which represents
a living organism under mutations. Furthermore, we changed the
genetic code used in the simulations to analyze its effect on
population stability.

\begin{figure}[!]
\begin{center}
\includegraphics[width=14cm]{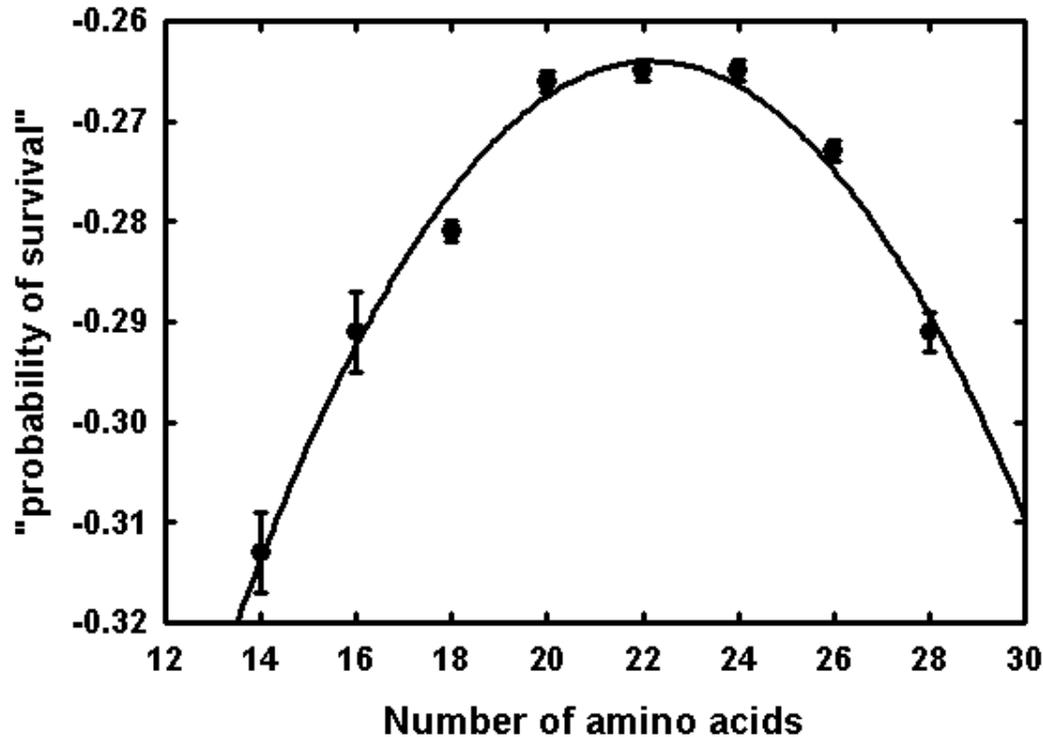}
\caption{``probabilities of survival" of different genetic code
tables. The fit to a parabola is just to giude the eye.}
\label{resultfig}
\end{center}
\end{figure}

All the results of different simulations are summarized in
Table \ref{results} and plotted in Figure \ref{resultfig}.

\begin{center}
\begin{table}[!]
\caption{Results of the simulations using different genetic code 
tables.} 
\label{results}
\begin{tabular}{|l|c|}
\hline Table Name & ``probability of survival" \\ 
\hline Nature & $-0.266 \pm 0.001$\\
\hline with 18 & $-0.281 \pm 0.001$\\
\hline with 16 & $-0.291 \pm 0.004$\\
\hline with 14 & $-0.313 \pm 0.004$\\
\hline with 18 (using MJ Matrix)& $-0.282 \pm 0.002$\\
\hline with 16 (using MJ Matrix)& $-0.287 \pm 0.003$\\
\hline with 14 (using MJ Matrix)& $-0.320 \pm 0.003$\\
\hline with 18 (using BLOSUM50 Matrix)& $-0.288 \pm 0.003$\\
\hline with 16 (using BLOSUM50 Matrix)& $-0.294 \pm 0.003$\\
\hline with 14 (using BLOSUM50 Matrix)& $-0.302 \pm 0.004$\\
\hline with 22 & $-0.265 \pm 0.001$\\
\hline with 24 & $-0.265 \pm 0.001$\\
\hline with 26 & $-0.273 \pm 0.001$\\
\hline with 28 & $-0.291 \pm 0.002$\\
\hline
\end{tabular}%
\end{table}
\end{center}

The results of shorter amino acid tables show that if we try to
conserve the information, the population which is represented by
less amino acids is affected more severely by mutations. However,
if we do not mind the information transferred by the gene (the
simulations with conserved genome length), the population is not
affected much.

Even if we use different reducing algorithms (MJ or BLOSUM50) for
genetic code, the population is affected by mutations more than
the population represented by the genetic code of Nature.

While we extend the amino acid table, we shorten the size of the
gene which means that we try to conserve the information. The
resistance of the population against mutations does not change
when the amino acid number is 22 or 24. However, after 24, the
resistance starts to decrease. 

The slopes of the simulations with 20, 22 and 24 amino acids are
very close. These results are along the same line with the results 
of the investigators from the Scripps Research Institute 
\cite{Anderson21930} and provides computational justification for their
belief that genetic codes with even more amino acids is feasible.
However the number of amino acids will be restricted to 20-22-24 if
we want the resulting life form to be resilient against mutations.

\section{ACKNOWLEDGEMENTS}
I am grateful to Dr. Isil Aksan Kurnaz and Dr. Muhittin Mungan for
their contributions on the model and the calculations.

\end{document}